\documentclass[singlecolumn,aps,superscriptaddress,eqsecnum,floatfix]{revtex4-1}
\usepackage{empheq}
\usepackage{amsmath}
\usepackage{amssymb}
\usepackage{slashed}
\usepackage{dsfont}
\usepackage{graphicx}
\usepackage{hyperref}
\usepackage{stackrel}
\usepackage{color}
\usepackage{comment}

\def \ee{\end{equation}}
\def \be{\begin{equation}}
\def \eea{\end{eqnarray}}
\def \bea{\begin{eqnarray}}

\begin{document}

\title{X17: A new force,\\ 
or evidence for  a hard $\gamma +\gamma$ process?}

\author{Benjamin Koch}
\email{bkoch@fis.puc.cl}
\affiliation{Pontificia Universidad Cat\'olica de Chile \\ Instituto de F\'isica, Pontificia Universidad Cat\'olica de Chile, \\
Casilla 306, Santiago, Chile}
\affiliation{Institut f\"ur Theoretische Physik,
 Technische Universit\"at Wien,
 Wiedner Hauptstrasse 8-10,
 A-1040 Vienna, Austria}

\begin{abstract}
It is investigated whether the ``X17 puzzle'' 
might be explained by
a nuclear decay chain
and a conversion of the two resulting highly energetic
$\gamma$s into an electron-positron pair. 
It is found that the
corresponding kinematics 
 fits perfectly to the experimental
result. Also the conversion rates of this process
are reasonable.
However, the assumed nuclear chain reaction
is not favored in the established nuclear models 
and no explanation for the isospin structure of the signal can be given.
Thus, it has to be concluded that the
process studied in this paper does not give a completely satisfying 
explanation of the ``X17 puzzle''.
\end{abstract}

{\color{blue}
\pacs{04.60.Gw,03.65.Pm}}
\maketitle


\tableofcontents

\section{Introduction}

\subsection{Inelastic $\gamma\gamma$ scattering}

The laws of classical physics do not allow for direct interaction
between two electromagnetic waves.
In contrast,  in quantum theory, 
such processes arise due to 
the contributions of virtual states.
Soon after the discovery of quantum electrodynamics, as
the  best theory describing  the quantum nature
of electromagnetic interactions, 
it became clear that the production of electron-positron pairs
from photon interactions is possible~\cite{Heisenberg:1934pza,Sauter:1931zz,Heisenberg:1935qt,Schwinger:1951nm}.
Over the years, inelastic light-light interactions
have been the  subject of numerous theoretical studies
and successful experimental tests~\cite{Burke:1997ew,Bamber:1999zt,Marklund:2006my,Abulencia:2006nb,Mourou:2006zz,Ruffini:2009hg,Schutzhold:2008pz,Bell:2008zzb,Aaltonen:2009kg,DiPiazza:2011tq,Aaboud:2017bwk}.

In this paper, it is pointed out that 
 the experimental results, known as the ``X17 puzzle''
might be added to this list.

\subsection{X17 experimental result}

Since 2015 the particle physics community faces 
an unexpected signature, which throughout this article will 
be called the ``X17 puzzle''.
This puzzle comes from an experiment which measured
the angular correlations and the kinematics of electron-positron
pairs, which were emitted during the relaxation of 
excited $^8Be^*$ nuclear states~\cite{Krasznahorkay:2015iga}.
In the angular distribution, they found an unexpected peak at a relatively
large opening angle ($\theta\approx 144^o$ ).
Taking the energy, the asymmetry, and the angle $\theta$ of the lepton pair on an event by event 
basis,  the corresponding invariant energy was calculated.
In this observable, a highly significant resonance at about $17$~MeV was discovered.
This result was  unexpected because it does not seem
to fit into the standard picture of nuclear and fundamental physics.

Recently, the same group announced a similar finding
from an experiment with excited $^4He^*$~\cite{Krasznahorkay:2019lyl}. 
This second excess, which appears at similar energy, 
but at significantly different angle, will not be discussed in this paper.

\subsection{X17  in the literature}

Internal formation of electron-positron pairs is a known effect in 
nuclear physics \cite{Horton:1948,Rose:1949zz,Hofmann:1990zq,Blinne:2016yzv}.
In this process, an emitted high energy $\gamma$ interacts non-perturbatively with 
the mostly Coulomb-like background field of the nucleus such that 
it converts into an electron-positron pair. This reaction is similar to
the Bethe-Heitler (BH) process, only that it differs in the initial state~\cite{Bethe:1934za}.
This type of process was taken into account in the background analysis
of the experimental studies. 
However, this background
is strongly increased towards small angles, and it shows
no peak in the resulting invariant mass spectrum.
For the case of $^8Be$ a more detailed study improved
the nuclear physics understanding of the observables by  considering
various additional effects such as the initial production process and
interference in an effective field theory
approach like the one used in \cite{Zhang:2017zap}.
It was shown that there are important additional effects that might 
reduce the significance of the reported observation, but it
was also shown that these effects can not account for the reported
experimental signature.

Given these inconsistencies with our current understanding of
nuclear physics,  different explanations were proposed.
An incomplete list of these explanations includes new forces~\cite{Feng:2016jff,Feng:2016ysn,Gu:2016ege,Fayet:2016nyc,Neves:2016nek,Kahn:2016vjr,Dror:2017nsg,Kozaczuk:2017per,DelleRose:2017xil,Pulice:2019xel},
new and dark matter~\cite{Alexander:2016aln,Ellwanger:2016wfe,Kitahara:2016zyb,Kozaczuk:2016nma,Jia:2016uxs,Chen:2016tdz,Liang:2016ffe,Battaglieri:2017aum,Dror:2017ehi,Krasznahorkay:2017gwn,DelleRose:2018pgm}, and axion-like particles \cite{Alves:2017avw,Bauer:2017ris}.
For a critical  revision of these ideas, see \cite{Fornal:2017msy}.
A common feature of these models is that they invoke a new mediator
 particle, which couples to the nuclear states and 
 leptonic states.

\subsection{X17 alternative}

The ``new particle'' hypothesis should
confirmed by complementary observables such the cross sections in
as direct production experiments $(e^+ + e^- \rightarrow X)$.
There are numerous such complementary experiments 
that are currently planned or operating~\cite{Denig:2016dgi,ATLAS:2016jza,Alikhanov:2017cpy,Chen:2016dhm,Kozhuharov:2017qjo,Nardi:2018cxi,Banerjee:2018vgk,Marsicano:2018krp,Taruggi:2018wha,Kozhuharov:2019kkf}.
However, it is possible that these experiments will not see any new fundamental resonance.

In this case, one has to consider alternative explanations for the X17 puzzle.
In this paper, one of the
intermediate nuclear states
in the decay chain of the excited $^8Be$ is studied.
These nuclear states would not show up in a direct production
experiment like $(e^+ + e^- \rightarrow X)$.
It will be discussed below whether such states
can produce a signal pattern that is very similar to the one
observed and reported in \cite{Krasznahorkay:2015iga}.

\section{Cascade trough a broad nuclear resonance}

\begin{figure}[h!]
\includegraphics[width=3\columnwidth/14]{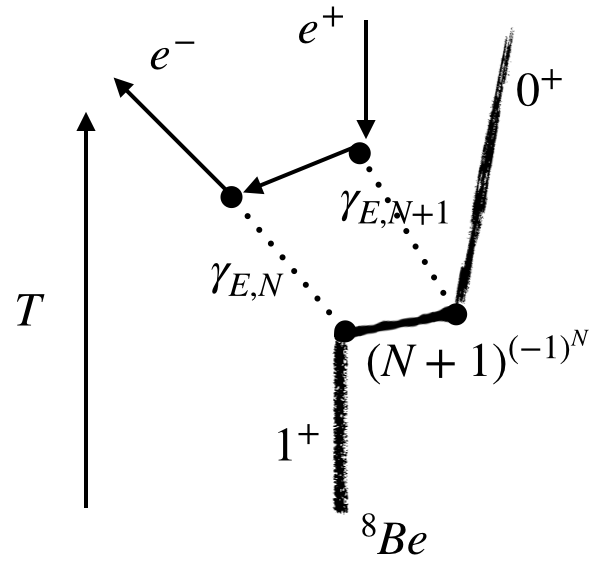}
\caption{
Modified Bethe-Heitler  transition~\cite{Bethe:1934za}
 for the $^8Be$ process.
 The process is also possible under a symmetric
interchange of electric and magnetic radiation $E \leftrightarrow  M$.
}
\label{fig:FBX17}
\end{figure}

It is proposed to study a modified Bethe-Heitler (MBH) process shown in figure \ref{fig:FBX17}.
Note that the original Bethe-Heitler transition is not 
a perturbative process \cite{Blinne:2016yzv}.
All calculations will be done in natural units, where $\hbar=c=1$.
In these transitions, angular momentum is changed by up to $N$ due to $N-$pole radiation.
Further, parity is changed by powers of $(-1)^N$ in electric,
and by powers of  $(-1)^{N+1}$ in magnetic
multipole radiation.
There are two main differences between this process and the usual Bethe-Heitler process.
The first one is that the
incoming $\gamma$ is provided
by excited $^8 Be$ and not by external radiation. 
The second difference is that the second $\gamma$ is provided
by another decay of the same nucleus and not by a background field.

It will now be investigated whether
the processes sketched in figure \ref{fig:FBX17}
can give experimental signatures like the ones
observed in  \cite{Krasznahorkay:2015iga}.
As starting point one can consider
 three ingredients:
\begin{itemize}
\item A broad intermediate state,
\item orientation of the nuclear multipole coefficients due to the emission of one of the $\gamma$s,
\item conservation of energy and momentum.
\end{itemize}
These ingredients will now shortly be explained in an on shell approximation of 
the process shown in  figure \ref{fig:FBX17}, before they are applied
to the  specific experiment.

\subsection{Broad intermediate state}

For the MBH process to have a non-vanishing probability for the conversion
$\gamma + \gamma \rightarrow e^+ + e^-$, the two $\gamma$s need to
conspire almost simultaneously in the region of the emitting nucleus.
The time delay $\Delta t$ between both $\gamma$  emissions will be given 
in terms of the width $\Gamma$ of the intermediate nuclear states
\be\label{delay}
\Delta t\sim \frac{1}{ \Gamma}.
\ee
Thus, a large width is needed to minimize 
this suppression of the probability of
$\gamma + \gamma \rightarrow e^+ + e^-$ conversion.

\subsection{Candidate state}

The X17 anomaly was  
measured in an experiment with excited $^8 Be$ states~\cite{Krasznahorkay:2015iga}.
The excited states had an overlap of the transitions $(J^\pi=1^+, T=1)\rightarrow (J^\pi=0^+, T=0)$ at $E_{13}=17.6$~MeV
and $(J^\pi=1^+, T=0)\rightarrow (J^\pi=0^+, T=0)$ at $E_{13}=18.2$~MeV.
Both processes will be considered below.

There are three known intermediate states between 
the excited initial sate(s)
 and the ground state \cite{Bronsti}.
These are a $(J^\pi=2^+, T=0)$ state at $\Delta M_{23}=3.03$~MeV above the ground state,
a $(J^\pi=2^+, T=0)$ state at $\Delta M_{23}=16.63$~MeV above the ground state,
and a $(J^\pi=4^+, T=0)$ state at $\Delta M_{23}=11.35$~MeV above the ground state.
Thus, there are, in principle, multiple transitions that could provide a contribution to the proposed MBH transition.
However, the broadest of these states is
$(J^\pi=4^+, T=0)$ at $\Delta M_{23}=11.35$~MeV with a width of $\Gamma_{4+}=3.5$~MeV~\cite{Bronsti,Fornal:2017msy}.
Thus, by virtue of (\ref{Area}), the isospin conserving
\be\label{BeDom}
{\bf 1^+}
\begin{array}{c}
{\tiny{
\Delta M_{12}=6.8~MeV}}\\
\xrightarrow{\hspace*{2cm}}\\
{\tiny{
\Delta N = 3,\Delta T=0}}
\end{array}
{\bf 4^+}
\begin{array}{c}
{\tiny{\Delta M_{23}=
11.35~MeV
}}\\
\xrightarrow{\hspace*{2cm}}\\
{\tiny{
\Delta N = 4,\Delta T=0}}
\end{array}
{\bf 0^+}
\ee
and the isospin violating transition
\be\label{BeDom2}
{\bf 1^+}
\begin{array}{c}
{\tiny{
\Delta M_{12}=6.3~MeV}}\\
\xrightarrow{\hspace*{2cm}}\\
{\tiny{
\Delta N = 3,\Delta T=1}}
\end{array}
{\bf 4^+}
\begin{array}{c}
{\tiny{\Delta M_{23}=
11.35~MeV
}}\\
\xrightarrow{\hspace*{2cm}}\\
{\tiny{
\Delta N = 4,\Delta T=0}}
\end{array}
{\bf 0^+}
\ee
will be the best candidate processes for this channel.

\subsection{Angular spectrum}

It will be shown how under relatively straight forward assumptions
the emission of two MeV gammas can come at a preferred relative angle $\theta_{rel}$.

As first approximation, one can describe the angular probabilities of the  electromagnetic emitted
with multipole radiation.
In the diagram~\ref{fig:FBX17}, one emitted $\gamma$ is from an electric N-pole with $|\Delta J|= N$ and $\Delta \pi=(-1)^N$.
Another emitted $\gamma$ is from electric $N+1$-pole radiation $|\Delta J|= N+1$ and $\Delta \pi=(-1)^{N+1}$.
It is clear that the direction
of highly energetic electric radiation in direction $\vec k_1$ will affect the orientation of
the multipole coefficients $a_{lm}$ of the remaining nucleus.
One can choose the coordinate system such that the  first
emission is aligned with the direction of the first emission $\vec k_1= |\vec k_1| \hat z$.
This emission will transfer a large amount of angular momentum to the remaining nucleus,
which then has multipole moments $a_{lm}$.
However, due to the directionality of the emission, 
the projection of this induced angular momentum onto the  $\hat z$ axis will be small,
or even zero.
This corresponds to the multipole coefficients  $a_{l0}\neq 0$. 
These multipole coefficients $a_{l0}$ will now be the source for the
subsequent emission.
The angular distribution of this ``following'' emission of multipole radiation will then be~\cite{Jackson:1999}
\be\label{JacksonFormula}
\frac{dP_{l0}}{d\theta} \sim \sin (\theta) |a_{l0}|^2 |\vec X_{l0}(\theta)|^2,
\ee
where $\vec X_{l0}(\theta)$ is proportional to the angular momentum operator acting
on a spherical harmonic function $Y_{l0}$ and where $\theta_{rel}$ is the angle between the two emissions.
 The most likely large relative angles with $\theta > 90^o$ between the two emissions are 
\be\label{thetam}
\theta_{rel}\pm \delta \theta_{rel}= \left\{
\begin{array}{ccc}
(144 \pm 14)^o   &\mbox{for}&  N=3\\
(152 \pm 11)^o   &\mbox{for}&  N=4\\
\dots
\end{array}
\right.
\ee
Here, $\Delta N=3,4$, were given because these
are the relevant quantum numbers involved in the $Be^8$ transition discussed in~(\ref{BeDom}).
In this process it is the transition from the intermediate $J^\pi=4^+$ to the $J^\pi=0^+$ ground state,
which is related to the maximum in the angular spectrum (\ref{thetam}).

\subsection{Kinematics}

The four momenta in figure \ref{fig:FBX17} are $p^\mu_1$
for $^8 Be$ in the initial state,
 $p^\mu_2$ for the intermediate nuclear state,
$p^\mu_3$ final nuclear state,
$k^\mu_1$ for the first $\gamma$,  $k_2^\mu$ for the second $\gamma$,
$q^\mu_1$ for the outgoing positron, $q^\mu_2$ for the intermediate lepton, and
 $q^\mu_3$ the outgoing electron.
The kinematics of the experiment allows for the approximations
\be\label{approx}
p_i^2\gg (p_i-p_j)|_{i\neq j}^2\gg m^2,
\ee
where $m$ is the electron mass and
where all particles are approximated to be on shell. 
For the nuclear part of the reaction, this approximation
is justified if the widths of the involved states do not overlap.
Within these approximations
the relative angle between the leptons 
and between the $\gamma$s is is the same.
The primary signal in~\cite{Krasznahorkay:2019lyl} was found
for small asymmetries $-0.5<y<0.5$, thus we will work in the center of mass frame
of a very heavy nucleus  with $y \approx 0$.
The approximations (\ref{approx}) are reasonable
since the nuclear masses are several GeV,  which is much larger than the
energies in the signal region which are of the order of $20$~MeV,
and the electron mass is $1/2$~MeV.
In these approximations, one finds
that the square of the invariant mass of the electron-positron pair in figure
\ref{fig:FBX17} is given by
\be\label{master}
m_X^2=(q_1+q_3)^2=4 (\Delta M_{12}) (E_{13}-\Delta M_{12}) \sin^2 \left( \frac{\theta}{2}\right),
\ee
where $\theta$ is the angle between the leptons and
$
\Delta M_{12}= \sqrt{p_1^2}-\sqrt{p_2^2}$
and
$
\Delta M_{23}= \sqrt{p_2^2}-\sqrt{p_3^2}=E_{13}-\Delta M_{12}$.
The derivation of this result can be found in the appendix II.A.
This result can get corrections when one considers virtual particles
in the intermediate states, which will be discussed in the next subsection.

\begin{figure}[h!]
\includegraphics[width=\columnwidth/2]{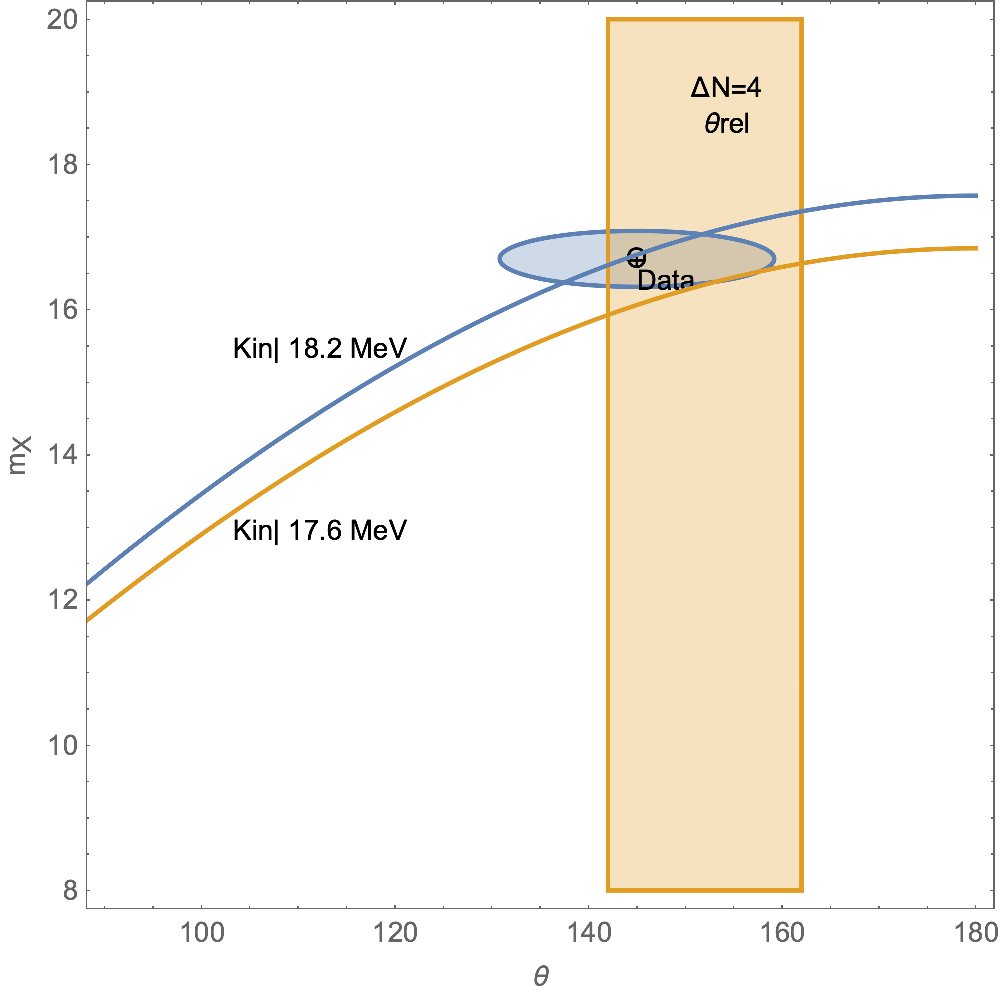}
\caption{$m_X$ as a function of the relative angle $\theta_{rel}$. 
The elliptic contour shows the experimental
result within the given uncertainties~\cite{Krasznahorkay:2015iga}. The blue curve
is obtained from the kinematic relation (\ref{master}) for the process (\ref{BeDom}).
 The orange curve
is obtained from the kinematic relation (\ref{master}) for the process (\ref{BeDom2}).
The vertical region is the angle $\theta_{rel}\pm \delta \theta_{rel}\pm \delta' \theta_{rel}$
compatible with the $\Delta N=4$ transition.
}
\label{fig:Bepred}
\end{figure}

With the energy differences for the above mentioned candidate state one can plot 
the leptonic invariant energy $m_X=\sqrt{(q_1+q_3)^2}$ of~(\ref{master})
as function of the relative angle $\theta_{rel}$. This relation 
for the processes (\ref{BeDom}) and (\ref{BeDom2})
is shown by the blue and orange curve
in figure \ref{fig:Bepred}.
One notes that both curves are compatible with the data.
The blue curve
crosses the observed data point almost perfectly,
supporting the kinematic relation (\ref{master}). This is a highly non trivial agreement because
there is no parameter tuned in these curves.

\subsection{Conversion probability of $\gamma + \gamma$ to $e^+ + e^-$}

\begin{figure}[h!]
\includegraphics[width=\columnwidth/2]{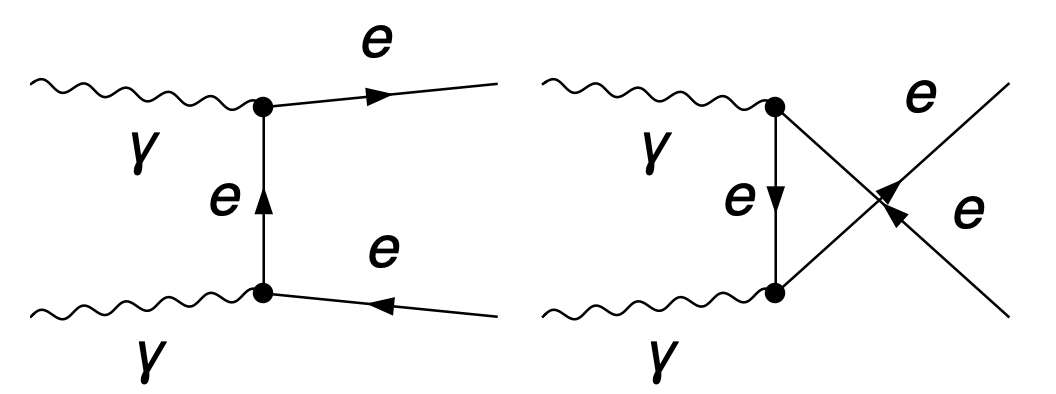}
\caption{
Feynman diagrams for the production of  $e^+$, $e^-$ from a $\gamma$ pair
by the interchange of a virtual positron/electron.
}
\label{fig:FProd}
\end{figure}

\begin{figure}[h!]
\includegraphics[width=\columnwidth/3]{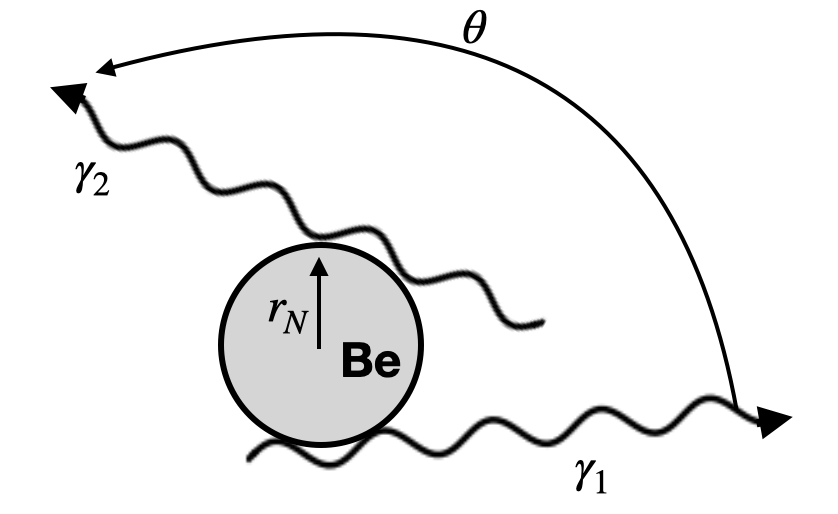}
\caption{
Geometric sketch of the emission of two $\gamma$ rays from one nucleus
}
\label{fig:impact}
\end{figure}

In this section the total conversion probability of the hard
 $\gamma + \gamma$ to $e^+ + e^-$ process will 
will be analyzed. 
The production process depicted in terms of the Feynman diagram shown in figure \ref{fig:FProd}.
The corresponding differential cross section in the center of mass frame, which was calculated with FeynArts \cite{Hahn:2000kx},
is
\be\label{dsigmadt}
\frac{d \sigma}{dt}=\frac{e^4}{32 \pi s^2}\frac{-m_e^4(3 s^2+4st+12 t^2)+me^2(s^3+2s^2t+8st^2+8t^3)+8tm_e^6-2m_e^8-t(s+t)(s^2+2st+2t^2)}
{ (t-m_e^2)^2(s-m_e^2+t)^2}
\ee
Here, the electric coupling $e$ is given in terms of the
 fine structure constant
 $\alpha_0=4 \pi/e^2 $.
Integrating this relation and inserting the kinematics of the candidate state
gives the total cross section 
\bea\label{sigma}\nonumber
\sigma&=&-\frac{e^4}{16 \pi s^3}\frac{(8m_e^6+3 m_e^2 s^2+s^3)\log(m_e^2/(s-m_e^2))+s(12 m_e^4 \log(s/m_e^2-1)-6 m_e^4+m_e^2 s +s^2)}{s-m_e^2}\\
&=&3.5 \cdot 10^{-6}\frac{1}{MeV^2}.
\eea
In order to translate this cross section to a conversion probability one needs to consider
the transversal area $A_t$ of the collision and a suppression factor $F$ due to the non-simultaneous emission of both $\gamma$ rays.
With this, the conversion probability can be estimated as
\be\label{conversion0}
p_{ \gamma+\gamma \rightarrow e^+ + e^-}\approx \frac{\sigma}{A_t}\cdot F.
\ee
The geometric picture which allows to estimate the transversal area $A_t$ is shown in figure \ref{fig:impact}.
One sees that both $\gamma$ rays originate from the same nucleus with radius $r_N$.
The transversal area $A_t$ is thus of the order of 
\be
A_t\approx \pi r_N^2.
\ee
The suppression factor comes $F$ from the fact that both $\gamma$ rays are not emitted 
simultaneously. The emission of the second $\gamma$ takes place with a delay of $\Delta t\sim 1/\Gamma$, where $\Gamma $ is the width  of the intermediate nuclear state.
The head of the wave train of the first gamma will thus have traveled a distance $l_\Gamma= c \Delta t\sim 1/ \Gamma$ before the second gamma appears.
If this distance is larger than the actual length of the wave trains $l_\gamma$, then the process
will be strongly suppressed. Both $\gamma$s will basically not meet each other.
For this suppression we assume an exponential behaviour
\be\label{Ffac}
F= \exp \left(-\frac{l_\Gamma}{l_\gamma}\right).
\ee
In a true momentum eigenstate $l_\gamma$ would be infinite and this suppression would
not take place. However, in order to be more conservative with the estimate of the conversion 
 factor we take the smallest possible length of the wave train, which would be given
by a single wave length. 
The energy of the photon $E_\gamma$ is inversely proportional to this length  $\lambda_\gamma \sim 1/E_\gamma$.
Thus, the suppression factor (\ref{Ffac}) reads
$
F\approx \exp \left(-\frac{E_\gamma}{\Gamma}\right).
$
With this, the conversion probability is
\be\label{conversion}
p_{ \gamma+\gamma \rightarrow e^+ + e^-}\approx \frac{\sigma}{\pi r_N^2}\cdot \exp{\left(-\frac{E_\gamma}{\Gamma}\right)}=7 \cdot 10^{-4},
\ee
where the nuclear radius of Beryllium was taken as $r_N=0.016/MeV$.
The result in (\ref{conversion}) is  below the conversion probability for the dominant 
background of internal pair creation $p_{bg}\approx 25 \cdot 10^{-4}$~\cite{Rose:1949zz}.
However, one should  take into account that the conversions (\ref{conversion}) would
be more concentrated in the signal region of large angles, while the background
conversions will be maximal at small relative angles and decreasing for
the large relative angles~\cite{Horton:1948}. Considering this, it is fair to say that
the conversion probabilities of the suggested MBH process
are in the ballpark of the background conversion probabilities $p_{bg}$.

In any case, the MBH process, if realized in nature, comes with a huge
enhancement of $\gamma$ pairs in the signal region. 

\subsection{Angular broadening due to off shell contributions of $\gamma + \gamma$ to $e^+ + e^-$}

The on-shell result (\ref{master}), 
relies strongly on
the approximation that the angle between $\vec k_1$ and $ \vec q_1$
vanishes $\theta_{rel} \approx 0$, just as the angle between
  $\vec k_2$ and $ \vec q_3$.
 The validity of this approximation can be checked by examining the differential
 cross section (\ref{dsigmadt}) in the laboratory frame.

From the explicit form of the amplitude
one realizes that 
the process is enhanced at very small
momentum interchange as assumed when deriving (\ref{master}).
Off-shell contributions and finite $m^2/q_i^0$ corrections
will induce a non-vanishing distribution of the relative angle $\theta_{rel} $.
This will, induce an additional width widening $\delta' \theta_{rel}$ of
a given maximal relative angle $\bar\theta_{rel}$ between $\vec q_1$ and $\vec q_3$.
It is an important consistency check that the combined angular 
width, arising from the initial emission $\delta \theta_{rel}$ and the
 kinematical widening  $\delta' \theta_{rel}$,
is in the ballpark of
the measured width (see \ref{widthBe}) of the distribution of the relative
lepton angles
\be\label{thetaInequality}
\sqrt{(2\delta \theta_{rel})^2+(2\delta' \theta_{rel})^2 }\approx \Gamma_\theta.
\ee

As sanity check one can contrast the observed $\Gamma_\theta$ with 
the additional widening of $\delta \theta_{rel}$ derived from 
the squared amplitude and see whether (\ref{thetaInequality}) is fulfilled.
The normalized squared amplitude in terms of the relative angle $\theta_{rel}$, as measured in
the center of mass frame of the lepton pair gives $\delta \theta_{rel}|_{CM}=3.5^o$.
Transforming this back to the laboratory frame one obtains a widening of $\delta' \theta_{rel}=\pm 8^o$,
which is sufficiently smaller than the $\Gamma_{\theta}=26^o$ obtained from (\ref{widthBe}).
Thus, the condition (\ref{thetaInequality}) holds.

There are further off-shell contributions from the other virtual particles, in  particular the
$\gamma$s . These are obtained from the box diagram in figure \ref{fig:FBX17}.
In order to perform this calculation, one would need an effective model
for the nuclear states, which goes beyond the scope of this paper.
Kinematically one can expect that the virtual $\gamma$s introduce effects of the same
order of magnitude  as (\ref{thetaInequality}).
The virtuality of the very heavy intermediate nuclear state will be negligible.


\subsection{Isospin}

The X-17 excess was found in the isospin conserving transition of the $18.15$~MeV initial state,
but it was not found in the isospin violating transition of the $17.64$~MeV initial state.
Thus, the isospin violating process (\ref{BeDom2}) must be suppressed with respect
to the isospin conserving process (\ref{BeDom}).
This is in agreement with the kinematic relation of figure \ref{fig:Bepred},
where one notices that  (\ref{BeDom}) gives a better match with the data than  (\ref{BeDom2}).
However, the kinematic arguments of the above discussion do not explain
such a selection. This is a strong argument against 
a solution of the ``X17 puzzle'' with a simple MBH process.

\begin{figure}[h!]
\includegraphics[width=3\columnwidth/12]{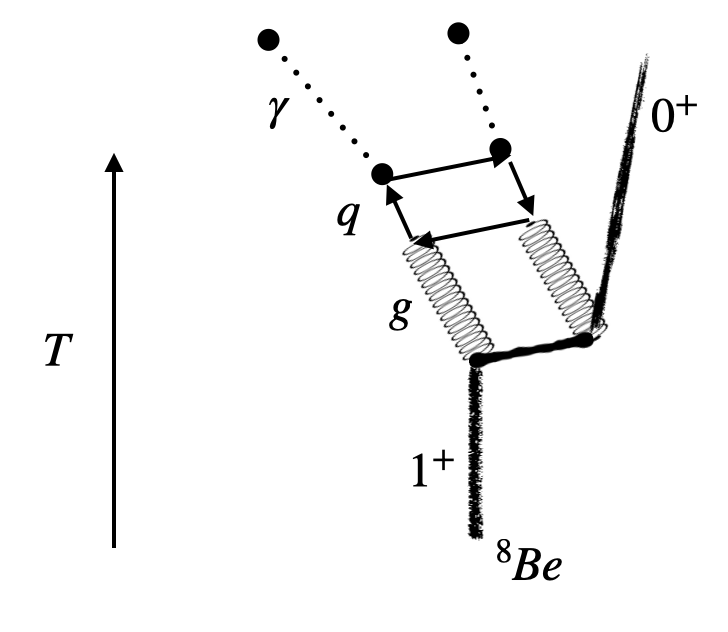}
\caption{
Strong subprocess, triggering the electromagnetic MBH 
transition of figure \ref{fig:FBX17}.
}
\label{fig:sub}
\end{figure}

One might only note that the strong coupling, in contrast to the electromagnetic coupling, conserves
isospin.
Thus, the suppression of a process like  (\ref{BeDom2}) could be a hint that the intermediate nuclear 
sate contains a substructure which involves a strongly interacting trigger of the
electromagnetic decay.
In this case, the reaction of the intermediate nuclear state in figure~\ref{fig:FBX17},
would arise from a strong subprocess, such as the one shown in figure \ref{fig:sub}.
This and other scenarios remain to be investigated, but they
 would not be in the spirit of this paper of seeking a solution 
of the puzzle within the established models of nuclear physics.

\subsection{Comments}

The above result deserves some comments.

\begin{itemize}
\item Complete model of $^8Be$:\\
In order to go beyond the presented relations one needs
a complete modeling of the excited $^8Be$ states and their transitions.
For example, the classical multipole formula (\ref{JacksonFormula}), 
is a first approximation to the underlying quantum
processes~\cite{Buck:1977zz,Langanke:1986zz,Baye:1992nhf,Wiringa:2000gb,Pieper:2004qw,Datar:2004sx,Pastore:2014oda,Hammer:2019poc}. 
There are also no strongly interacting subprocess, like the one shown in figure \ref{fig:sub}, considered.
Further, from the most common nuclear model one would expect that most intermediate states at
$11.35$~MeV would decay into another broad intermediate state at $3.5$~MeV.
It is planned improve the understanding of 
these two points by using an effective field theory 
description for the nuclear states in the spirit of~\cite{Zhang:2017zap,Fornal:2017msy,Hammer:2019poc}.

\item  Possible peaks at small angles:\\
The classical multipole formula allows also for peaks at small angles.
These are likely to be invisible due to the large background at small angles~(see  \ref{fig:pAnBe}).

\item Initial state at $16.7$~MeV:\\
In~\cite{Krasznahorkay:2015iga} an overlaping initial state at $16.7$~MeV is
reported. As shown in figure~\ref{fig:Bepred}, the kinematic relation (\ref{master}) for this initial state would
also be in agreement with the observed signal. However, as reported in~\cite{Krasznahorkay:2015iga},
no unusual excess is observed in the angular distribution of the lepton
pairs originating from the $16.7$~MeV initial state. A possible explanation for this
could be that, in addition to the with of the intermediate state $\Gamma_{int}$,
the entire MBH reaction needs
to occur at a small time window $\Delta t\sim\frac{1}{\Gamma_{ini}}+\frac{1}{\Gamma_{int}}$.
Since the width of the $16.7$~MeV initial state is by a factor of 14 smaller than the width
of the  width of the $18.2$~MeV initial state, this could lead to a suppression of the MBH transition
for the former.

\item A smoking gun:\\
Very few highly energetic $\gamma$s are converted
to electron-positron pairs through the MBH process.
The conversion probability can be estimated from (\ref{conversion}).
This turns out to be if the order of $7 \cdot 10^{-4}$.
Thus, most $\gamma$s leave the nucleus without conversion.
Measuring their angular distribution 
would provide a smoking gun signal for the MBH process.

\item
There is another experimental result with 
a peak at about $17~MeV$~\cite{Krasznahorkay:2019lyl}. 
It would be nice to offer a simultaneous explanation for 
both of these experiments. This is, not possible with the
MBH idea, since there is no known intermediate state with the
desired properties of excited Helium.
However, it has been questioned that both results originate from the same 
physical process~\cite{Siegel:2019} because they appear at different relative angle $\theta_{rel}$.

\end{itemize}

\section{Conclusion}

\begin{figure}[h!]
\includegraphics[width=8\columnwidth/12]{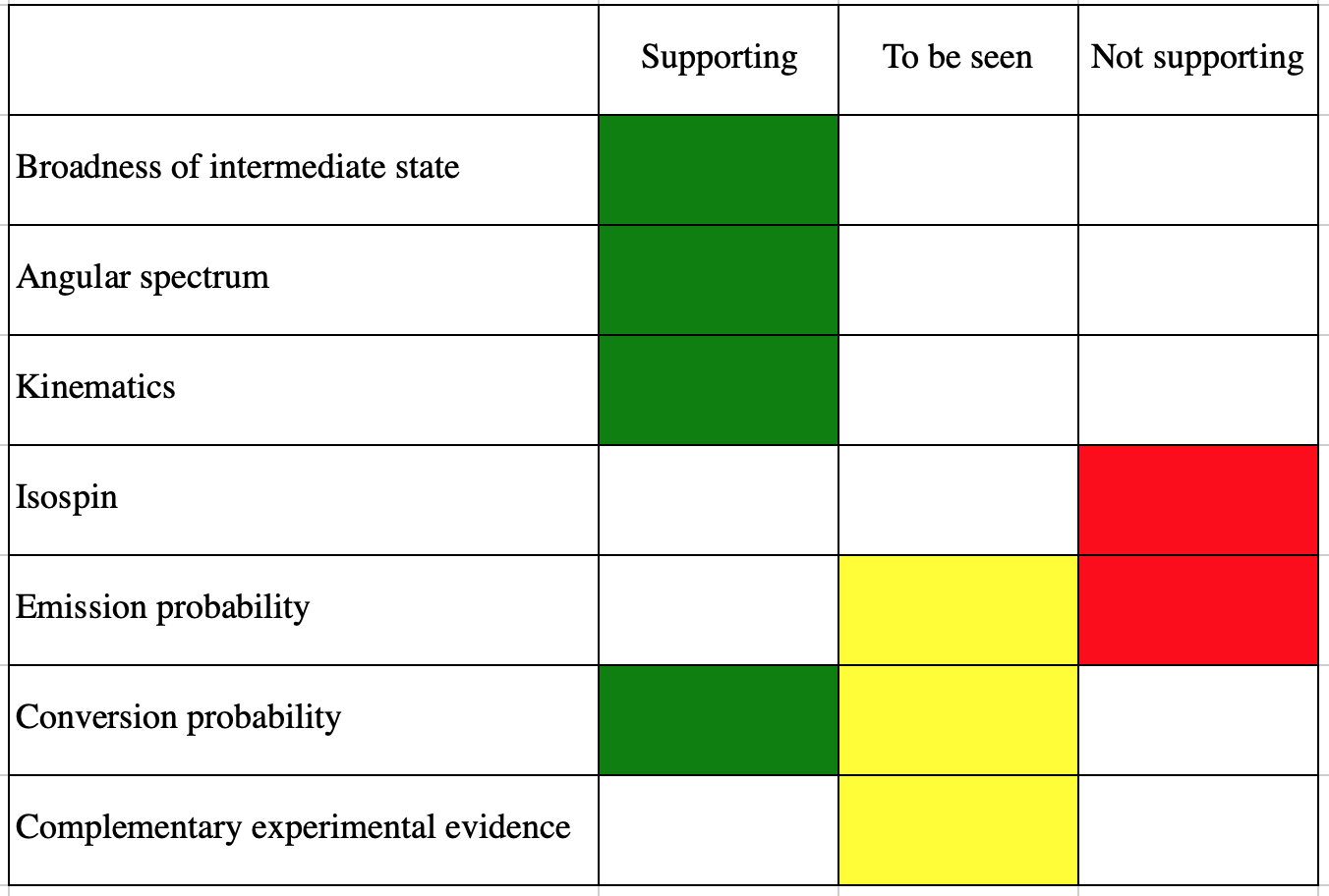}
\caption{
Summary of the aspects which favor or disfavor the MBH hypothesis of this paper.
While the existence of a broad intermediate state, the expected angular distribution, 
and the kinematic relations support this idea, the
isospin blindness, the emission probabilities, and the conversion probabilities
disfavor such an explanation.  Complementary experiments, such as the measurement of coincidence rates
in the $\gamma + \gamma$ channel would certainly give valuable
information.
}
\label{fig:ConclTable}
\end{figure}

This paper explored a new possibility
of explaining the observed X17 anomaly
reported in \cite{Krasznahorkay:2015iga}.
The study focused on a modified Bethe-Heitler transition, depicted in figure \ref{fig:FBX17}.
For this process, one needs to consider a broad intermediate nuclear resonance
and standard conservation of energy and momentum~(\ref{master}). 
The intermediate resonance is provided by the $^8Be^*(J^\pi=4^+, \,T=0)$ state
with a large width of $\Gamma=3.5$~MeV
at $11.35$~MeV above the ground state. Even though one finds an almost perfect
agreement with the kinematic relation~(\ref{master}),
there sere several aspects which make it unlikely that the ``X17'' anomaly
can be explained by rates arising from the simple processes mentioned above.
This is the  main result of this paper,  which is  summarized in the table in
figure \ref{fig:ConclTable}

The idea of this work was to find
a convincing explanation of the ``X17'' results with a modified Bethe-Heitler process.
However, the answer turned out to be most likely negative, unless 
one can come up with some effect which considerably enhances the corresponding
emission and transition probabilities.
A simple direct test the hypothesis of this paper could be made
by measuring the angular distribution in the two $\gamma$ final state.

\newpage

\section{Acknowledgements}
 Many thanks to J.~Schaffner-Bielich, H.~Stoecker, C.~Greiner, A.~Hoang, H.~Skarke, P.~Arias,  V. Datar, 
  and C.~Diaz for
 helpful remarks.
 Thanks to several colleagues from nuclear physics, for insisting on the issue
 with emission and conversion probabilites.
This work was supported by Fondecyt 1181694.

\section*{Appendix I: Data fit}

\begin{figure}[h!]
\includegraphics[width=\columnwidth/12*5]{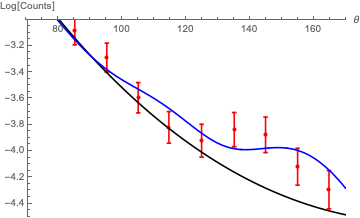}
\includegraphics[width=\columnwidth/12*5]{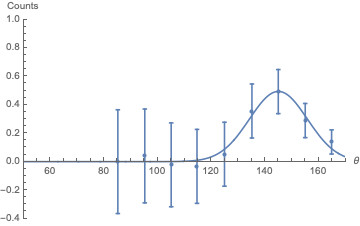}
\caption{
Angular distribution of  the  lepton  final 
state~\cite{Krasznahorkay:2019lyl}.
The upper figure shows the signal in red, the interpolated
background in black, and a widened $\Delta  N=4$ angular distribution~(\ref{JacksonFormula})
added to the background in blue. 
The lower figure shows signal minus background
and the Gaussian fit.
The error bars in the upper figure are smoothly approximated to~\cite{Krasznahorkay:2019lyl}, 
in the lower figure they are scaled due to the change in the signal to background
ratio.
}
\label{fig:pAnBe}
\end{figure}

By using \cite{WebDigit}
one can extract  information on the angular distributions of the electron-positron pairs
reported in~\cite{Krasznahorkay:2015iga}, as shown in figure~ \ref{fig:pAnBe}.

This distribution is fitted by a Gaussian \\
$
N_\theta=0.49 e^{-\frac{(\theta-\theta_{rel})^2}{2 \sigma_\theta}},
$
peaked at the angle
$
\theta_{rel}=145^o,
$
with a halfwidth 
\be\label{widthBe}
\Gamma_\theta=2 \sqrt{2 \ln (2)} \sigma_\theta=26 ^o.
\ee
With a statistical error of $18 \%$.


\section*{Appendix II: Derivations}

\subsection*{A: Kinematic relation \ref{master}}

Given the initial nucleus $p_1^\mu=(M_1,0,0,0)$
and the first photon four momentum $k_1^\mu=(k_1^0,k_1^0,0,0)$,
the four momentum of the intermediate nuclear state is
\be\label{p2}
p_2^\mu=p_1^\mu-k_1^\mu.
\ee
Here, the $\hat z$ axes was aligned with the direction of the first photon.
The invariant mass square of the intermediate state is 
\be\label{p22}
p_2^\mu p_{2\;\mu}=(M_1-\Delta M_{12})^2,
\ee
where $\Delta M_{12}$ is the mass difference between the 
initial and the intermediate nuclear state.
Inserting (\ref{p2}) in (\ref{p22}) one can solve for the first photon energy
\be
k_1^0=\frac{2 M_1 \Delta M_{12}-\Delta M_{12}^2}{2 M_1}.
\ee
The second photon can pick up an additional spatial direction
which can, for convenience be chosen as the $\hat y$ direction
$k_2^\mu=\left(k_2^0,k_2^0\cdot \cos(\theta),k_2^0\cdot \sin(\theta),0\right)$.
The four momentum of the final nuclear state is
\be\label{p3}
p_3^\mu=p_2^\mu-k_2^\mu.
\ee
The invariant mass square of the final nuclear state is
\be\label{p32}
p_3^\mu p_{3\;\mu}=(M_1-\Delta M_{12}-\Delta M_{23})^2,
\ee
where $\Delta M_{23}$ is the rest mass difference between the intermediate
and the final nuclear states.
Inserting (\ref{p3}) in (\ref{p32}) one can solve for the first photon energy
\be
k_1^0=\frac{2 M_1 \Delta M_{22}(2M1-2\Delta M_{12}+\Delta M_{12})}{2 M_1^2-2M_1 \Delta M_{12}+\Delta M_{12}^2+\Delta M_{12}(2M_1-\Delta M_{12})\cdot \cos (\theta)}.
\ee
With this, the invariant mass square of the final photon state is
\bea
m_X^2&=&(k_1+k_2)_\mu (k_1+k_2)^\mu\\
&=&\frac{2 \Delta M_{12} \Delta M_{23}(2M_1-\Delta M_{12})(2M_1-(2\Delta M_{12}+\Delta M_{23})) \sin^2 (\theta/2)}{-2M1^2+2M_1\Delta M_{12}-\Delta M_{12}^2+\Delta M_{12}(-2M_1 + \Delta M_{12})\cos (\theta)}.
\eea
For $M_1\gg \Delta M_{ij}$ this simplifies to
\be
m_X^2\approx 4 \Delta M_{12}\Delta M_{23} \sin^2\left(\frac{\theta}{2}\right),
\ee
as given in (\ref{master}).

\newpage


\begin{thebibliography}{99}


\bibitem{Heisenberg:1934pza} 
  W.~Heisenberg,
  Z.\ Phys.\  {\bf 90}, no. 3-4, 209 (1934)
  Erratum: [Z.\ Phys.\  {\bf 92}, no. 9-10, 692 (1934)].
  doi:10.1007/BF01340782, 10.1007/978-3-642-70078-1 6, 10.1007/978-3-642-70078-1 8, 10.1007/BF01333516

\bibitem{Sauter:1931zz} 
  F.~Sauter,
  Z.\ Phys.\  {\bf 69}, 742 (1931).
  doi:10.1007/BF01339461

\bibitem{Heisenberg:1935qt} 
  W.~Heisenberg and H.~Euler,
  Z.\ Phys.\  {\bf 98}, no. 11-12, 714 (1936)
  doi:10.1007/BF01343663, 10.1007/978-3-642-70078-1 9
  [physics/0605038].

\bibitem{Schwinger:1951nm} 
  J.~S.~Schwinger,
  Phys.\ Rev.\  {\bf 82}, 664 (1951).
  doi:10.1103/PhysRev.82.664



\bibitem{Burke:1997ew} 
  D.~L.~Burke {\it et al.},
  Phys.\ Rev.\ Lett.\  {\bf 79}, 1626 (1997).
  doi:10.1103/PhysRevLett.79.1626

\bibitem{Bamber:1999zt} 
  C.~Bamber {\it et al.},
  Phys.\ Rev.\ D {\bf 60}, 092004 (1999).
  doi:10.1103/PhysRevD.60.092004

\bibitem{Mourou:2006zz} 
  G.~A.~Mourou, T.~Tajima and S.~V.~Bulanov,
  Rev.\ Mod.\ Phys.\  {\bf 78}, 309 (2006).
  doi:10.1103/RevModPhys.78.309

\bibitem{Marklund:2006my} 
  M.~Marklund and P.~K.~Shukla,
  Rev.\ Mod.\ Phys.\  {\bf 78}, 591 (2006)
  doi:10.1103/RevModPhys.78.591
  [hep-ph/0602123].
 
\bibitem{Abulencia:2006nb} 
  A.~Abulencia {\it et al.} [CDF Collaboration],
  Phys.\ Rev.\ Lett.\  {\bf 98}, 112001 (2007)
  doi:10.1103/PhysRevLett.98.112001
  [hep-ex/0611040].
 
\bibitem{Bell:2008zzb} 
  A.~R.~Bell and J.~G.~Kirk,
  Phys.\ Rev.\ Lett.\  {\bf 101}, 200403 (2008).
  doi:10.1103/PhysRevLett.101.200403
  
\bibitem{Schutzhold:2008pz} 
  R.~Schutzhold, H.~Gies and G.~Dunne,
  Phys.\ Rev.\ Lett.\  {\bf 101}, 130404 (2008)
  doi:10.1103/PhysRevLett.101.130404
  [arXiv:0807.0754 [hep-th]].
  
\bibitem{Ruffini:2009hg} 
  R.~Ruffini, G.~Vereshchagin and S.~S.~Xue,
  Phys.\ Rept.\  {\bf 487}, 1 (2010)
  doi:10.1016/j.physrep.2009.10.004
  [arXiv:0910.0974 [astro-ph.HE]].
  
\bibitem{Aaltonen:2009kg} 
  T.~Aaltonen {\it et al.} [CDF Collaboration],
  Phys.\ Rev.\ Lett.\  {\bf 102}, 242001 (2009)
  doi:10.1103/PhysRevLett.102.242001
  [arXiv:0902.1271 [hep-ex]].
  
  
\bibitem{DiPiazza:2011tq} 
  A.~Di Piazza, C.~Muller, K.~Z.~Hatsagortsyan and C.~H.~Keitel,
  Rev.\ Mod.\ Phys.\  {\bf 84}, 1177 (2012)
  doi:10.1103/RevModPhys.84.1177
  [arXiv:1111.3886 [hep-ph]].

\bibitem{Aaboud:2017bwk} 
  M.~Aaboud {\it et al.} [ATLAS Collaboration],
  Nature Phys.\  {\bf 13}, no. 9, 852 (2017)
  doi:10.1038/nphys4208
  [arXiv:1702.01625 [hep-ex]].
  


\bibitem{Krasznahorkay:2015iga} 
  A.~J.~Krasznahorkay {\it et al.},
  Phys.\ Rev.\ Lett.\  {\bf 116}, no. 4, 042501 (2016)
  doi:10.1103/PhysRevLett.116.042501
  [arXiv:1504.01527 [nucl-ex]].

\bibitem{Krasznahorkay:2019lyl} 
  A.~J.~Krasznahorkay {\it et al.},
  arXiv:1910.10459 [nucl-ex].


\bibitem{Bethe:1934za} 
  H.~Bethe and W.~Heitler,
  Proc.\ Roy.\ Soc.\ Lond.\ A {\bf 146}, 83 (1934).
  doi:10.1098/rspa.1934.0140

\bibitem{Horton:1948}
G.~K.~Horton,
Proc\  Phys.\ Soc.\ {\bf{61}} 3, 296
Corrections to 1948 
Proc.\ Phys.\ Soc. 60 457,. (1948)

\bibitem{Rose:1949zz} 
  M.~E.~Rose,
  Phys.\ Rev.\  {\bf 76}, 678 (1949)
  Erratum: [Phys.\ Rev.\  {\bf 78}, 184 (1950)].
  doi:10.1103/PhysRev.76.678, 10.1103/PhysRev.78.184

\bibitem{Blinne:2016yzv} 
  A.~Blinne,
  arXiv:1701.00743 [physics.plasm-ph].
  
\bibitem{Hofmann:1990zq} 
  C.~Hofmann, J.~Reinhardt, W.~Greiner, P.~Schluter and G.~Soff,
  Phys.\ Rev.\ C {\bf 42}, 2632 (1990).
  doi:10.1103/PhysRevC.42.2632

\bibitem{Zhang:2017zap} 
  X.~Zhang and G.~A.~Miller,
  Phys.\ Lett.\ B {\bf 773}, 159 (2017)
  doi:10.1016/j.physletb.2017.08.013
  [arXiv:1703.04588 [nucl-th]].


\bibitem{Feng:2016jff} 
  J.~L.~Feng, B.~Fornal, I.~Galon, S.~Gardner, J.~Smolinsky, T.~M.~P.~Tait and P.~Tanedo,
  Phys.\ Rev.\ Lett.\  {\bf 117}, no. 7, 071803 (2016)
  doi:10.1103/PhysRevLett.117.071803
  [arXiv:1604.07411 [hep-ph]].


\bibitem{Feng:2016ysn} 
  J.~L.~Feng, B.~Fornal, I.~Galon, S.~Gardner, J.~Smolinsky, T.~M.~P.~Tait and P.~Tanedo,
  Phys.\ Rev.\ D {\bf 95}, no. 3, 035017 (2017)
  doi:10.1103/PhysRevD.95.035017
  [arXiv:1608.03591 [hep-ph]].


\bibitem{Gu:2016ege} 
  P.~H.~Gu and X.~G.~He,
  Nucl.\ Phys.\ B {\bf 919}, 209 (2017)
  doi:10.1016/j.nuclphysb.2017.03.023
  [arXiv:1606.05171 [hep-ph]].


\bibitem{Fayet:2016nyc} 
  P.~Fayet,
  Eur.\ Phys.\ J.\ C {\bf 77}, no. 1, 53 (2017)
  doi:10.1140/epjc/s10052-016-4568-9
  [arXiv:1611.05357 [hep-ph]].

\bibitem{Neves:2016nek} 
  M.~J.~Neves and J.~A.~Helayel Neto,
  Annalen Phys.\  {\bf 530}, no. 3, 1700112 (2018)
  doi:10.1002/andp.201700112
  [arXiv:1609.08471 [hep-ph]].

\bibitem{Kahn:2016vjr} 
  Y.~Kahn, G.~Krnjaic, S.~Mishra-Sharma and T.~M.~P.~Tait,
  JHEP {\bf 1705}, 002 (2017)
  doi:10.1007/JHEP05(2017)002
  [arXiv:1609.09072 [hep-ph]].

\bibitem{Dror:2017nsg} 
  J.~A.~Dror, R.~Lasenby and M.~Pospelov,
  Phys.\ Rev.\ D {\bf 96}, no. 7, 075036 (2017)
  doi:10.1103/PhysRevD.96.075036
  [arXiv:1707.01503 [hep-ph]].

\bibitem{Kozaczuk:2017per} 
  J.~Kozaczuk,
  Phys.\ Rev.\ D {\bf 97}, no. 1, 015014 (2018)
  doi:10.1103/PhysRevD.97.015014
  [arXiv:1708.06349 [hep-ph]].

\bibitem{DelleRose:2017xil} 
  L.~Delle Rose, S.~Khalil and S.~Moretti,
  Phys.\ Rev.\ D {\bf 96}, no. 11, 115024 (2017)
  doi:10.1103/PhysRevD.96.115024
  [arXiv:1704.03436 [hep-ph]].

\bibitem{Pulice:2019xel} 
  B.~Pulice,
  arXiv:1911.10482 [hep-ph].


\bibitem{Alexander:2016aln} 
  J.~Alexander {\it et al.},
  arXiv:1608.08632 [hep-ph].

\bibitem{Ellwanger:2016wfe} 
  U.~Ellwanger and S.~Moretti,
  JHEP {\bf 1611}, 039 (2016)
  doi:10.1007/JHEP11(2016)039
  [arXiv:1609.01669 [hep-ph]].

\bibitem{Kitahara:2016zyb} 
  T.~Kitahara and Y.~Yamamoto,
  Phys.\ Rev.\ D {\bf 95}, no. 1, 015008 (2017)
  doi:10.1103/PhysRevD.95.015008
  [arXiv:1609.01605 [hep-ph]].

\bibitem{Kozaczuk:2016nma} 
  J.~Kozaczuk, D.~E.~Morrissey and S.~R.~Stroberg,
  Phys.\ Rev.\ D {\bf 95}, no. 11, 115024 (2017)
  doi:10.1103/PhysRevD.95.115024
  [arXiv:1612.01525 [hep-ph]].


\bibitem{Jia:2016uxs} 
  L.~B.~Jia and X.~Q.~Li,
  Eur.\ Phys.\ J.\ C {\bf 76}, no. 12, 706 (2016)
  doi:10.1140/epjc/s10052-016-4561-3
  [arXiv:1608.05443 [hep-ph]].

\bibitem{Chen:2016tdz} 
  C.~S.~Chen, G.~L.~Lin, Y.~H.~Lin and F.~Xu,
  Int.\ J.\ Mod.\ Phys.\ A {\bf 32}, no. 31, 1750178 (2017)
  doi:10.1142/S0217751X17501780
  [arXiv:1609.07198 [hep-ph]].

\bibitem{Liang:2016ffe} 
  Y.~Liang, L.~B.~Chen and C.~F.~Qiao,
  Chin.\ Phys.\ C {\bf 41}, no. 6, 063105 (2017)
  doi:10.1088/1674-1137/41/6/063105
  [arXiv:1607.08309 [hep-ph]].

\bibitem{Battaglieri:2017aum} 
  M.~Battaglieri {\it et al.},
  arXiv:1707.04591 [hep-ph].

\bibitem{Dror:2017ehi} 
  J.~A.~Dror, R.~Lasenby and M.~Pospelov,
  Phys.\ Rev.\ Lett.\  {\bf 119}, no. 14, 141803 (2017)
  doi:10.1103/PhysRevLett.119.141803
  [arXiv:1705.06726 [hep-ph]].

\bibitem{Krasznahorkay:2017gwn} 
  A.~J.~Krasznahorkay {\it et al.},
  EPJ Web Conf.\  {\bf 142}, 01019 (2017).
  doi:10.1051/epjconf/201714201019


\bibitem{DelleRose:2018pgm} 
  L.~Delle Rose, S.~Khalil, S.~J.~D.~King and S.~Moretti,
  Front.\ in Phys.\  {\bf 7}, 73 (2019)
  doi:10.3389/fphy.2019.00073
  [arXiv:1812.05497 [hep-ph]].


\bibitem{Bauer:2017ris} 
  M.~Bauer, M.~Neubert and A.~Thamm,
  JHEP {\bf 1712}, 044 (2017)
  doi:10.1007/JHEP12(2017)044
  [arXiv:1708.00443 [hep-ph]].

\bibitem{Alves:2017avw} 
  D.~S.~M.~Alves and N.~Weiner,
  JHEP {\bf 1807}, 092 (2018)
  doi:10.1007/JHEP07(2018)092
  [arXiv:1710.03764 [hep-ph]].


\bibitem{Fornal:2017msy} 
  B.~Fornal,
  Int.\ J.\ Mod.\ Phys.\ A {\bf 32}, 1730020 (2017)
  doi:10.1142/S0217751X17300204
  [arXiv:1707.09749 [hep-ph]].




\bibitem{Denig:2016dgi} 
  A.~Denig,
  EPJ Web Conf.\  {\bf 130}, 01005 (2016).
  doi:10.1051/epjconf/201613001005

\bibitem{ATLAS:2016jza} 
  The ATLAS collaboration [ATLAS Collaboration],
  ATLAS-CONF-2016-042.

\bibitem{Alikhanov:2017cpy} 
  I.~Alikhanov and E.~A.~Paschos,
  Phys.\ Rev.\ D {\bf 97}, no. 11, 115004 (2018)
  doi:10.1103/PhysRevD.97.115004
  [arXiv:1710.10131 [hep-ph]].

\bibitem{Chen:2016dhm} 
  L.~B.~Chen, Y.~Liang and C.~F.~Qiao,
  arXiv:1607.03970 [hep-ph].

\bibitem{Kozhuharov:2017qjo} 
  V.~Kozhuharov,
  EPJ Web Conf.\  {\bf 142}, 01018 (2017).
  doi:10.1051/epjconf/201714201018

\bibitem{Nardi:2018cxi} 
  E.~Nardi, C.~D.~R.~Carvajal, A.~Ghoshal, D.~Meloni and M.~Raggi,
  Phys.\ Rev.\ D {\bf 97}, no. 9, 095004 (2018)
  doi:10.1103/PhysRevD.97.095004
  [arXiv:1802.04756 [hep-ph]].

\bibitem{Banerjee:2018vgk} 
  D.~Banerjee {\it et al.} [NA64 Collaboration],
  Phys.\ Rev.\ Lett.\  {\bf 120}, no. 23, 231802 (2018)
  doi:10.1103/PhysRevLett.120.231802
  [arXiv:1803.07748 [hep-ex]].

\bibitem{Marsicano:2018krp} 
  L.~Marsicano {\it et al.},
  Phys.\ Rev.\ D {\bf 98}, no. 1, 015031 (2018)
  doi:10.1103/PhysRevD.98.015031
  [arXiv:1802.03794 [hep-ex]].

\bibitem{Taruggi:2018wha} 
  C.~Taruggi [PADME Collaboration],
  ``Searching for dark Photons with the PADME Experiment,''

\bibitem{Kozhuharov:2019kkf} 
  V.~Kozhuharov,
  EPJ Web Conf.\  {\bf 212}, 06001 (2019).
  doi:10.1051/epjconf/201921206001

\bibitem{Jackson:1999}
J.~D.~Jackson
http://cdsweb.cern.ch/record/490457 (1999).


\bibitem{WebDigit}
Ankit Rohatgi;
 https://automeris.io/WebPlotDigitizer.

\bibitem{Bronsti}
Sukhoruchkin, \& Soroko(2008)]{2008LanB..19C...55S} Sukhoruchkin, S.~I., \& Soroko, Z.~N.\ 2008, Landolt B\&ouml;rnstein, 19C, 55. 

\bibitem{Hahn:2000kx}
T.~Hahn,
Comput. Phys. Commun. \textbf{140} (2001), 418-431
doi:10.1016/S0010-4655(01)00290-9
[arXiv:hep-ph/0012260 [hep-ph]].

\bibitem{Buck:1977zz} 
  B.~Buck, H.~Friedrich and C.~Wheatley,
  Nucl.\ Phys.\ A {\bf 275}, 246 (1977).
  doi:10.1016/0375-9474(77)90287-1

\bibitem{Langanke:1986zz} 
  K.~Langanke and C.~Rolfs,
  Phys.\ Rev.\ C {\bf 33}, 790 (1986).
  doi:10.1103/PhysRevC.33.790

\bibitem{Baye:1992nhf}
D.~Baye, P.~Descouvemont and M.~Kruglanski,
Nucl. Phys. A \textbf{550}, 250-262 (1992)
doi:10.1016/0375-9474(92)90682-A

\bibitem{Wiringa:2000gb} 
  R.~B.~Wiringa, S.~C.~Pieper, J.~Carlson and V.~R.~Pandharipande,
  Phys.\ Rev.\ C {\bf 62}, 014001 (2000)
  doi:10.1103/PhysRevC.62.014001
  [nucl-th/0002022].

\bibitem{Pieper:2004qw} 
  S.~C.~Pieper, R.~B.~Wiringa and J.~Carlson,
  Phys.\ Rev.\ C {\bf 70}, 054325 (2004)
  doi:10.1103/PhysRevC.70.054325
  [nucl-th/0409012].

\bibitem{Datar:2004sx} 
  V.~M.~Datar, S.~Kumar, D.~R.~Chakrabarty, V.~Nanal, E.~T.~Mirgule, A.~Mitra and H.~H.~Oza,
  Phys.\ Rev.\ Lett.\  {\bf 94}, 122502 (2005)
  Erratum: [Phys.\ Rev.\ Lett.\  {\bf 94}, 139902 (2005)]
  doi:10.1103/PhysRevLett.94.122502, 10.1103/PhysRevLett.94.139902
  [nucl-ex/0409025].

\bibitem{Pastore:2014oda} 
  S.~Pastore, R.~B.~Wiringa, S.~C.~Pieper and R.~Schiavilla,
  Phys.\ Rev.\ C {\bf 90}, no. 2, 024321 (2014)
  doi:10.1103/PhysRevC.90.024321
  [arXiv:1406.2343 [nucl-th]].

\bibitem{Hammer:2019poc} 
  H.-W.~Hammer, S.~Konig and U.~van Kolck,
  arXiv:1906.12122 [nucl-th].



\bibitem{Siegel:2019}
E.~Siegel, Forbes science section, 26.11.2019.


\end{thebibliography}
\end{document}